\documentclass[twocolumn]{aa}
\usepackage{graphicx}
\begin{document}

\title{HD\,77407 and GJ\,577: two new young stellar binaries}

\subtitle{detected with the Calar Alto Adaptive Optics system ALFA}

\author{M. Mugrauer \inst{1,2,3}
\and R. Neuh\"auser \inst{1,2,3} \and E.W. Guenther \inst{3,4} \and A.P. Hatzes \inst{4}\and
\linebreak N. Hu\'elamo \inst{3,5} \and M. Fern\'andez \inst{3,6} \and M. Ammler \inst{1,3} \and
J. Retzlaff \inst{2,3}, B. K\"onig \inst{2}, D. Charbonneau \inst{7},
R. Jayawardhana \inst{8} \and W. Brandner \inst{9}}

\offprints{Markus Mugrauer, markus@astro.uni-jena.de}

\institute{Astrophysikalisches Institut, Universit\"at Jena, Schillerg\"a{\ss}chen 2-3, 07745 Jena, Germany \\
\and MPI f\"ur extraterrestrische Physik, Giessenbachstra{\ss}e, 85748 Garching, Germany \\
\and Visiting Astronomer, German-Spanish Astronomical Centre, Calar Alto \\
\and Th\"uringer Landessternwarte Tautenburg, Sternwarte 5, 07778 Tautenburg, Germany \\
\and European Southern Observatory,  Alonso de Cordova 3107, Casilla 19001, Santiago, Chile \\
\and Instituto de Astrof\'{\i}sica de Andaluc\'{\i}a, CSIC, Apdo. Correos
3004, 18080 Granada, Spain \\
\and California Institute of Technology, mail stop 105-24, 1200 E. California Blvd., Pasadena CA 91125, USA \\
\and Department of Astronomy, University of Michigan, Ann Arbor MI 48109, USA \\
\and MPI f\"ur Astronomie, K\"onigstuhl 17, 69117 Heidelberg, Germany }

\date{Received 16 May 2003 / Accepted 8 December 2003}

\abstract{We present the first results from our search for close stellar and sub-stellar companions
to young nearby stars on the northern sky. Our infrared imaging observations are obtained with the
3.5\,m Calar Alto telescope and the AO system ALFA. With two epoch observations which were
separated by about one year, we found two co-moving companion candidates, one close to HD\,77407
and one close to GJ\,577. For the companion candidate near GJ\,577, we obtained an optical spectrum
showing spectral type M4.5; this candidate is a bound low-mass stellar companion confirmed by both
proper motion and spectroscopy. We estimate the masses for HD\,77407\,B and GJ\,577\,B to be $\sim
0.3$ to $0.5\,M_{\odot}$ and $\sim 0.16$ to $0.2\,M_{\odot}$, respectively. Compared to Siess al.
(2000) models, each of the two pairs appears co-eval with HD\,77407\,A,B being 10 to 40\,Myrs and
GJ\,577\,A,B being $\ge$ 100\,Myrs old. We also took multi-epoch high-resolution spectra of
HD\,77407 to search for sub-stellar companions, but did not find any with $3\,M_{Jup}$ as upper
mass ($m\sin\,i$) limit (for up to 4 year orbits); however, we detected a long-term radial velocity
trend in HD\,77407\,A, consistent with a $\sim 0.3$\,M$_{\odot}$ companion at $\sim 50$\,AU
separation, i.e. the one detected by the imaging. Hence, HD\,77407\,B is confirmed to be a bound
companion to HD\,77407\,A. We also present limits for undetected, but detectable companions using a
deep image of HD\,77407\,A and B, also observed with the Keck NIRC2 AO system; any brown dwarfs
were detectable outside of 0.5\,arcsec (17\,AU at HD\,77407), giant planets with masses from $\sim
6.5$ to 12\,M$\rm_{Jup}$ were detectable at $\ge 1.5$\,arcsec.

}

\maketitle

\section{Introduction}

Most nearby stars are quite old, so that close sub-stellar companions are too faint to be detected
directly. If we consider young stars, their companions are also young and therefore self-luminous
due to accretion and contraction, see e.g. Wuchterl \& Tscharnuter (2003), who consider objects
even younger than in our observations; other teams (e.g. Baraffe et al. 1998, Burrows et al. 1997)
also show quantitatively how sub-stellar objects get fainter when they get older. However they use
arbitrary initial conditions, so that their models should not be used for objects younger than
$\sim 10$\,Myrs (Baraffe et al. 2002). If such young companions are also nearby, they should be
well separated from their primaries. Hence, young nearby stars are most suitable for direct imaging
of sub-stellar companions, see e.g. Jayawardhana \& Greene (2001) for a comprehensive overview.
There are $\sim 200$ stars known with ages from $\sim 1$ to 100\,Myrs within $\sim 100$\,pc (e.g.
Montes et al. 2001b, Wichmann et al. 2003), about one third of them being in the northern sky. Most
of our targets still show lithium absorption and/or Ca II, H and K emission; some of them also show
H$\alpha$, and/or are classified as members of young associations by Montes et al. (2001b).

The direct detection of close sub-stellar companions is currently less difficult in the near
infrared, because in these wavelengths the brightness difference between companion and primary star
is low and detectors work well; in the thermal infrared, the brightness difference is even lower,
but the detectors are not yet as sensitive as in JHK. Nevertheless these close companions are much
fainter than their host star. The application of an adaptive optics system increases the resolution
to the diffraction limit and advances the dynamic range so that the detection of these faint
objects becomes feasible.

The determination of the spectral type and hence companionship of faint companion candidates using
JHK colors is difficult, because such objects are too faint and located in the PSF wing of their
host star, so that colors cannot be measured well. To solve simultaneously both extinction and
spectral type, three-band imaging would be indispensable. Nevertheless one needs a further
observation to confirm the proper motion of a possible companion. Finally four images are necessary
which makes such a photometric search inefficient. On the other hand, an astrometric survey (in one
band) only needs two observations (1st and 2nd epoch), i.e. allows one to study many more targets
with a minimum of observation time.

Most of the nearby stars have high proper motion and therefore they are well suited for an
astrometrical survey. In this technique each star with at least one faint object nearby is observed
in two epochs. Companion and primary star show the same motion relative to non-moving background
stars. The proper motion of our target stars is high enough so that an epoch difference of one year
is sufficient in most cases to find co-moving companions.

For a co-moving companion (candidate), spectroscopic confirmation is always necessary, either by
taking a spectrum of the companion (showing its late spectral type) and/or by taking spectra of the
primary (showing its secular acceleration due to the companion). We are searching for sub-stellar
companions also on the southern sky, with speckle and normal imaging at the ESO NTT, and now also
with AO (NAOS-CONICA at the ESO VLT). Four sub-stellar companions to young (nearby) stars have been
confirmed by both proper motion and spectroscopy: The $\sim 12$\,Myrs young TWA-5 B (Lowrance et
al. 1999, Neuh\"auser et al. 2000), the $\sim 300$ Myrs old Gl\,569\,B,C (Mart\'in et al. 2000,
Lane et al. 2001), the $\sim 35$\,Myrs young HR\,7239\,B (Lowrance et al. 2000, Guenther et al.
2001), and the $\sim 300$\,Myrs old HD\,130948\,B,C (Potter et al. 2002, Goto et al. 2002).

Here, we present some first results from our ongoing imaging program at the Calar Alto observatory.
We present the instrument and data reduction in Sect.\,2, the astrometric results in Sect.\,3,
photometry in Sect.\,4, and spectroscopy in Sect.\,5 \& 6. Finally, in Sect.\,7, we present the H-R
diagram to determine masses and ages of the new companions and discuss our results.

\section{Imaging observations and data reduction}

The observations were done in the H-Band (1.6\,$\mu$m) using the 3.5\,m telescope on Calar Alto
observatory in Spain. The telescope was equipped with the adaptive optics system ALFA (for Adaptive
optics with a Laser guide star For Astronomy, Glindemann et al. 2000), used here without laser
guide star. The IR detector used was $\Omega$-Cass, a $1024\times1024$ HgTeCd-detector with a pixel
resolution of 0.077\,$''$ per pixel. As individual integration time, we used 0.842\,s, i.e. as
short as possible in order to avoid saturation on the bright primary stars. In order to reach a
high sensitivity (i.e. limiting magnitude for faint companions), the total integration time was
almost 20\,min. Therefore, many short integrated images had to be superimposed.

The jitter technique was used to take into account the high IR sky background and to observe both
sky and target star in each frame. After sky subtraction, each image was flat fielded with a mean
sky flat image, created out of several sky images taken in twilight. Finally all reduced images
were shifted and combined to the result frame.

Up to now we have observed $\sim 80$ stars out of our northern sample of $\sim 100$ stars at least
once. We observed 12 stars with companion candidates in two different epochs. 22 more stars with at
least one faint object nearby still require a 2nd epoch image. In the following, we will present
two companions found among these 12 stars observed twice. The remaining data will be presented
later elsewhere.

In addition to Calar Alto, we have also observed a few stars with the Keck~2 telescope on Mauna
Kea, Hawai'i. We observed HD\,77407 on 28 Feb 2002 using the Keck~2 telescope with NIRC2 and AO
using an H-band filter and the 400\,mas diameter coronograph. NIRC2 has a pixel scale of
9.942$\pm$0.500\,mas/pixel (NIRC2 team, Campbell, priv. comm.). The total integration time for the
observations of HD\,77407 was almost 7\,min (23$\times$(100$\times$0.182\,s). We performed the data
reduction using the reduction software MIDAS provided by ESO for flat fielding, background
subtraction, and adding of the images.

The 400\,mas coronograph has a throughput of about $\sim$ 0.1\,\% determined by us by comparing the
companion HD\,77407\,B outside the coronograph with the primary star A located behind the
coronograph. Note that we use the 400\,mas coronograph, different from the 300\,mas coronograph
used by K\"onig et al. (2002) for another observation obtained in the same night.

\section{Astrometric results}

The first step in the astrometric analysis of the images is the measurement of the pixel scale and
the orientation of the images. In order to determine these important values, we need to observe
visual binary stars with well-known separations and orientations (position angles), e.g. from
Hipparcos, namely HD\,79210, HD\,82159, HD\,108574, HD\,112733, HD\,218738, HIP\,63322. The result
of the calibration for the different epochs is shown in Table\,\ref{Tabelle}.

The second step is the position measurement of the target star and all faint objects in the images,
all of which may be regarded as companion candidates. The position measurement is done with
ESO-MIDAS using gaussian centering.

We have observed 12 stars with companion candidates twice, i.e. at two different epochs separated
by about one year (see below for details). In the case that a companion candidate is a non-moving
background object, we should see only the (known and fast) motion of the star, i.e. the separation
changes from 1st to 2nd epoch. This motion includes the proper motion and the parallactic motion.
Moreover, for close companions, orbital motion may be detectable. If the separations measured in
1st and 2nd epoch in both right ascension and declination are identical (within the errors and
taken into account possible orbital motion), then a faint object can be regarded as co-moving
companion -- unless the proper motion of the primary star is too slow: The motion of the primary
star between the two images (according to its proper motion and the epoch difference) should be
significantly larger than the astrometric precision achieved, in order to obtain significant
astrometric results. A follow-up spectrum is always useful to confirm the companionship.

\begin{table} [htb]
\centering \caption{Pixel scale and image orientation of Omega-Cass for the different observing
runs. North is shifted by the orientation angle given below from the top of the image to right.}
\label{Tabelle}
\begin{tabular}{c|c|c}
epoch      & pixel scale $[$mas$]$ & orientation \\ \hline
01 Nov 2001 & 77.40$\pm$0.20 & 11.80$\pm$0.30$^{\circ}$  \\
26 Apr 2002 & 77.60$\pm$0.30 & 22.27$\pm$0.05$^{\circ}$  \\
23 Dec 2002 & 77.46$\pm$0.05 & 18.60$\pm$0.10$^{\circ}$  \\
\end{tabular}
\end{table}

\subsection{HD\,77407}

We have included HD\,77407 in our sample of young nearby stars after the Cool Stars Workshop in
1999 on Tenerife, where it was listed as a young star in the poster by Wichmann \& Schmitt (2001),
see also Montes et al. (2001a) and Wichmann et al. (2003). In our own optical spectrum (see
Sect.\,6) we measure EW(Li)=170\,m\AA~i.e. the star is clearly young.

\begin{figure} [htb] \resizebox{\hsize}{!}{\includegraphics{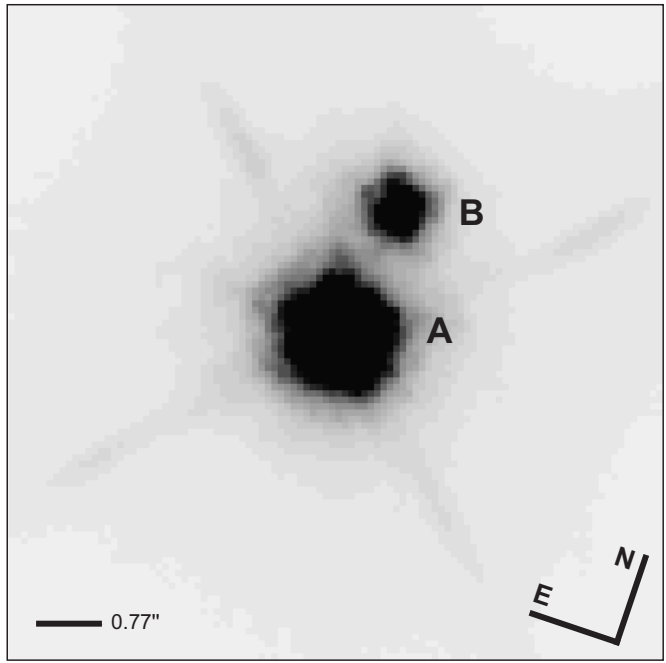}} \caption{HD\,77407: H-Band
image taken in epoch Dec 2002 with the 3.5\,m Calar Alto + ALFA + $\Omega$-Cass with a total
integration time of 16.5\,min} \label{hd77}
\end{figure}

HD\,77407 is located at a distance of 30.1$\pm$0.8\,pc. The proper motion of this star is -86.26$
\pm$1.24\,mas in right ascension and -168.79$\pm$0.59\,mas in declination.

The ALFA image is shown in Fig.\,\ref{hd77}. There is a close object (B) north of the primary
target star (A) at a separation of only $1.689\pm0.005$\,$''$ with a position angle of
353.7$\pm$0.5\,$^{\circ}$. The star was observed in epoch Nov 2001 and Dec 2002. The separation
between B and A for the two epochs is shown in Table\,\ref{tabellehd77}.

If this candidate is a non-moving background star, we should see only the proper and parallactic
motion of the star between the two observing epochs. This motion can be calculated using Hipparcos
data. Table\,\ref{tabellermhd77} shows the measured relative motion between the star and the
companion candidate which can be computed with the values from Table\,\ref{tabellehd77}. The
relative motion between the star and its companion candidate -- as observed -- is significantly
smaller than expected for a non-moving background star. Therefore, it appears likely that object B
is a real companion of HD\,77407. We do not include the Keck image in this analysis, because it has
a different pixel scale. Precise relative astrometry should be done only by using the same
telescope and instrument.

There is some relative motion between the two objects larger than 5\,$\sigma$ in declination. How
can we explain this? A separation of 1.689$\pm$0.005\,$''$ corresponds to a projected separation of
$50\pm2$\,AU at a distance of 30.1$\pm$0.8\,pc. We assume the mass of the barycenter to be $\sim
1$\,M$_{\sun}$ (this is reasonable because the primary has spectral type G0 and lies in an H-R
diagram close to the main sequence). Using Keplers's third law, we can estimate the orbital motion
of the companion. The predicted separation is 31.5\,mas at the second epoch observations. This is
consistent with the observed relative motion between HD\,77407\,A and B (41$\pm$11\,mas, Table\,3).
According to Fig.\,2, we can clearly distinguish between a non-moving background object and a
co-moving companion (allowing for orbital motion). Hence, the pair appears to be bound and we see
first hints for orbital motion.

\begin{table} [htb]
\centering \caption[]{Separation between HD\,77407\,A and B} \label{tabellehd77}
\begin{tabular}{c|c|c}
epoch      &  RA $[$ arcsec $]$ & DEC $[$ arcsec $]$ \\
\hline
Nov 2001 & 0.197$\pm$0.009 & -1.644$\pm$0.005 \\
Dec 2002 & 0.174$\pm$0.004 & -1.680$\pm$0.002 \\
\end{tabular}
\end{table}

\begin{table} [htb]
\centering \caption{Relative motion between HD\,77407\,A and B} \label{tabellermhd77}
\begin{tabular}{c|c|c}
&  RA $[$ arcsec $]$ & DEC $[$ arcsec $]$ \\
\hline
result of astrometry & -0.023$\pm$0.010 & -0.034$\pm$0.005 \\
(observed) & & \\ \hline
if B is non-moving & -0.109  & -0.180 \\
(calculated) & & \\
\end{tabular}
\end{table}

\begin{figure} [htb] \resizebox{\hsize}{!}{\includegraphics{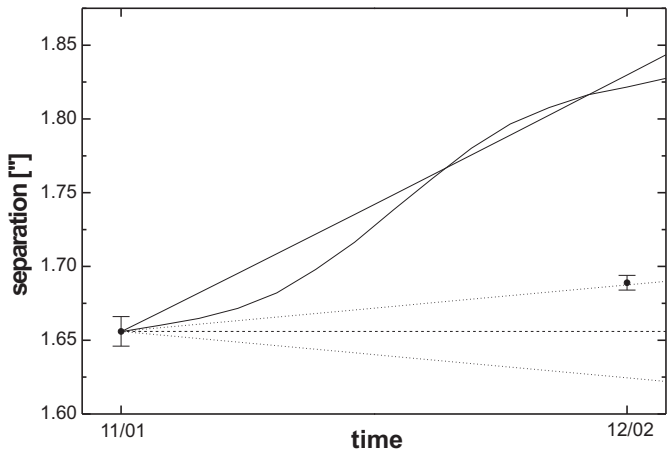}} \caption{Separation of
HD\,77407B from its host star at the two observations. The straight solid line is the expected
separation if B is a non-moving background star; the curved solid line takes into account
parallactic motion of the primary. The dotted line ($\pm$ the calculated orbital motion of B) is
expected for a co-moving companion. The 2nd epoch observation is $27\,\sigma$ deviant from the
background hypothesis and fully consistent with being a co-moving companion.} \label{hd77sep}
\end{figure}

\subsection{GJ\,577}

GJ\,577 is included in our sample since 2001, because it is listed as young star in Montes et al.
(2001a, 2001b).

\begin{figure} [htb]\resizebox{\hsize}{!}{\includegraphics{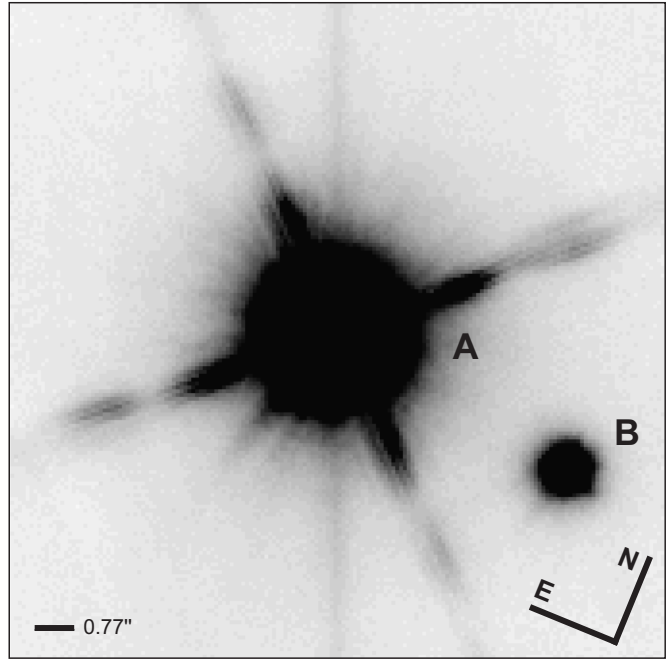}} \caption{GJ\,577: H-Band image
taken in epoch April 2002 with the 3.5\,m Calar Alto + ALFA + $\Omega$-Cass with a total integration
time of 17.9\,min} \label{gj577}
\end{figure}

GJ\,577 is located at a distance of 44.3$\pm$1.3\,pc. Its proper motion is -121.47$\pm$0.73\,mas in
right ascension and $112.21\pm0.63$\,mas in declination. The ALFA image is shown in
Fig.\,\ref{gj577}. There is an object (B) to the west of the star (A) in a separation of
5.39$\pm$0.02\,$''$ with a position angle of 260.9$\pm$0.2\,$^{\circ}$. The star was observed in
epoch April and Dec 2002. The separation between B and GJ\,577 is shown in
Table\,\ref{tabellegj577}. We can again calculate the relative motion between the star and its
companion candidate. With the Hipparcos data, we determine the motion of the star between the two
epochs. The results are shown in Table\,\ref{tabellermgj577}. The computed relative motion for a
non-moving background star is listed below. In both axis the relative motion between the two
objects is negligible if we consider the uncertainty in the astrometry. In declination the observed
relative movement is 10 times smaller as the value expected for a non-moving background object.
Therefore, we can conclude that object B is a companion of GJ\,577. A separation of 5.39\,$''$ (at
44.3$\pm$1.3\,pc) corresponds to a semi-major axis of 239$\pm7$\,AU. Therefore, the orbital motion
of the companion can be calculated with the third Kepler law. We again assume a mass of barycenter
of $1\,M_{\sun}$ because the primary has a spectral type G5 and lies close to the main sequence in
an H-R diagram. We then expect an orbital motion of about 7\,mas within the time passed between the
first and second observation in April and December 2002. The prediction is consistent with the
measured orbital motion of the primary and secondary, see Fig.\,4.

\begin{table} [htb]
\centering \caption[]{Separation between GJ\,577\,A and B} \label{tabellegj577}
\begin{tabular}{c|c|c}
epoch      &  RA $[$ arcsec $]$ & DEC $[$ arcsec $]$ \\
\hline
April 2002 & 5.332$\pm$0.018 & 0.865$\pm$0.012 \\
Dec 2002 & 5.325$\pm$0.003 & 0.836$\pm$0.009 \\
\end{tabular}
\end{table}

\begin{table} [htb]
\centering \caption{Relative motion between GJ\,577\,A and B} \label{tabellermgj577}
\begin{tabular}{c|c|c}
&  RA $[$ arcsec $]$ & DEC $[$ arcsec $]$ \\
\hline
result of astrometry & -0.007$\pm$0.018 & -0.029$\pm$0.015 \\
(observed) & & \\ \hline
if B is non-moving & -0.071  & 0.036 \\
(calculated) & & \\
\end{tabular}
\end{table}

\begin{figure} [htb] \resizebox{\hsize}{!}{\includegraphics{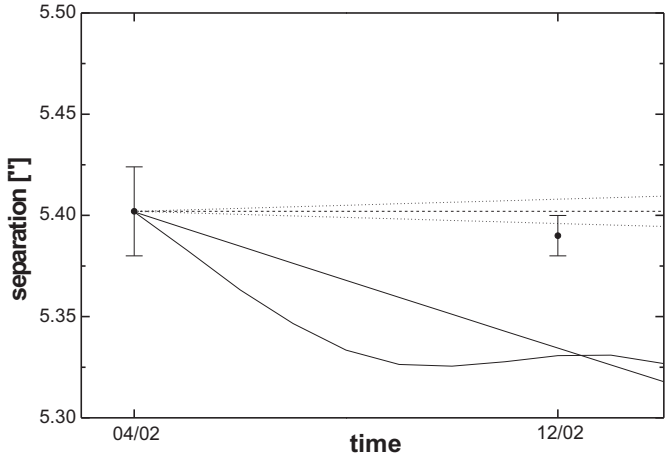}} \caption{Separation of
GJ\,577\,B from its host star at the two observations. The straight solid line is the expected
separation if B is a non-moving background star; the curved solid line takes into account
parallactic motion of the primary. The dotted line ($\pm$ orbital motion as expected) is expected,
if B is co-moving with A.} \label{gj577sep}
\end{figure}

\section{Photometry}

Due to the small field of view of an adaptive optical system, it is not possible to find many 2MASS
reference stars for relative photometry. Therefore, we determine the magnitude difference between
the star and the companion. With the H-band magnitude for star A as published in 2MASS, and having
assured that the primary stars are not saturated in our images, we can derive the H-band magnitude
of the companions. The results are shown in Table\,\ref{photometry}.

\begin{table} [htb]
\centering \caption{Photometry of HD\,77407\,A,B and GJ\,577\,A,B apparent magnitudes (from 2MASS
for the primaries A, from our images for the companions B by differential photometry).}
\label{photometry}
\begin{tabular}{l|c|c}
Name & m$_{H}$(A) & m$_{H}$(B)\\
\hline
HD\,77407& ~\,5.54 $\pm$ 0.03 & ~\,7.68 $\pm$ 0.07\\
GJ\,577  & ~\,6.89 $\pm$ 0.02 & ~\,10.84 $\pm$ 0.05\\
\end{tabular}
\end{table}

From the known spectral types of the primaries and, hence, their expected intrinsic JHK colors and
their published JHK magnitudes (from 2MASS), we find that interstellar absorption is negligible, as
expected for nearby stars. Hence, the absolute H-band magnitudes for the companions are 5.3$\pm$0.1
for HD\,77407\,B and 7.6$\pm$0.1 for GJ\,577\,B. We do not use the Keck image of HD\,77407 for
photometry here, because star A is located behind a (semi-transparent) coronograph and star B is
located partially behind one of the spiders.

\section{Spectroscopy of GJ\,577\,B}

Spectra of GJ\,577\,A and B were taken with the MOSCA faint-object spectrograph on the  Calar Alto
3.5\,m telescope at the end of Dec 2002. We use the green500 grism which covers the wavelength
range from 4200 to 8200\,\AA . The dispersion is about 2.9\,\AA~per pixel with Site-CCD which has
2048x4096 15\,$\mu m$ pixel. We used a slit-width of one arcsec which gives a resolving power
$\lambda/\Delta \lambda \simeq 700$. In order to calibrate the relative fluxes in the spectra, we
also observed Feige\,56. Standard IRAF routines were used to flat field, wavelength and
flux-calibrate the spectra using frames taken with the standard flat field and Hg-Ar and Ne lamps.

\begin{figure} [htb] \resizebox{\hsize }{!}{\includegraphics[angle=270]{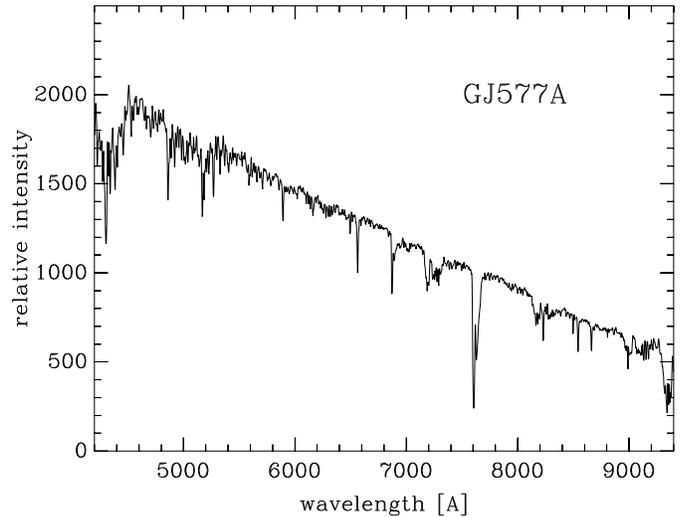}}
\caption{Optical spectrum of GJ\,577\,A, the primary, G5.}
\end{figure}

\begin{figure} [htb] \resizebox{\hsize}{!}{\includegraphics[angle=270]{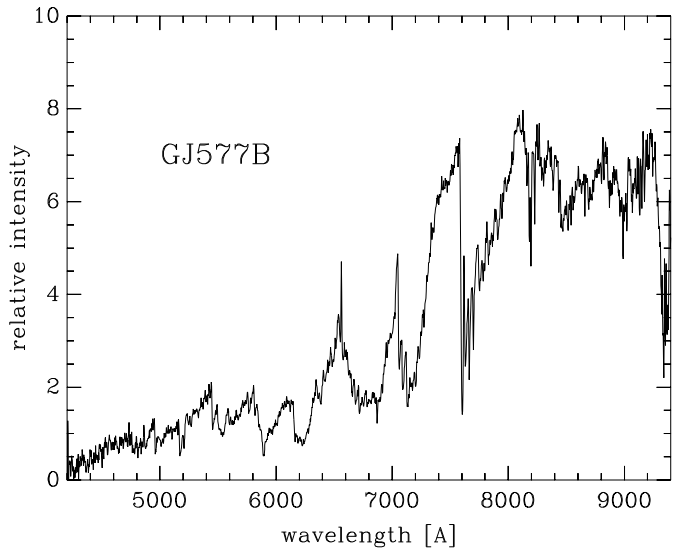}}
\caption{Optical spectrum of GJ\,577 B, the companion, M4.5, showing strong H$_{\alpha}$ emission,
deep TiO and VO molecular absorbtion bands and red continuum.}
\end{figure}

Fig.\,5 shows the resulting spectrum of GJ\,577\,A and Fig.\,6 the spectrum of GJ\,577\,B. As
pointed out by Reid et al. (1995), the ratio of the fluxes in the 7042 to 7046\,\AA -band versus
the flux in the 7126 to 7135\,\AA~band can be used for determining the temperatures of very
late-type objects. We derive TiO-2, TiO-4 and TiO-5 values of 0.508, 0.587 and 0.340 respectively.
Using these values and the information given in Reid et al. (1995) and Hawley et al. (1996), we
derive spectral types in the range M4 to M5. We also used the PC3 (1.02) and PC6 (2.05)
coefficients from Mart\'in et al. (1999) yielding M3.1 and M3 to M4, respectively; for the PC6
index, used for L-dwarfs only in Mart\'in et al. (1999), we used data of early M dwarfs given in
that paper and a linear regression to early M. The spectrum also shows that the H$\alpha$-line is
in emission. The equivalent width is -4.7\,\AA , which implies that $log(H_{\alpha}/L_{bol})$ is
about -4.0. These values are fairly typical for an object of this spectral type, and we thus
conclude that the level of chromospheric activity of GJ\,577\,B is normal. The equivalent width of
the Na I doublet around 8190\AA~is 5.8\AA , partly affected by telluric absorption. It appears
deeper (stronger) than in $\sim 10$\,Myrs young M-types objects of the TW Hya association
(Neuh\"auser et al. 2000), pointing to an older age for GJ\,577 B compared to TW Hya stars.

Given the observed magnitude difference between primary and secondary, and the known spectral type
G5 for the primary, the spectral type M4 to M5 for the fainter object is as expected for a bound
companion, i.e. being at the same age, distance, and metallicity.

\section{Spectroscopy of HD\,77407\,A}

We have started a program to search for close, low-mass companions around young nearby stars using
precise stellar radial velocity (RV) measurements taken with the 2m-Alfred-Jensch telescope of the
Th\"uringer Landessternwarte Tautenburg (TLS). Young stars have been largely excluded from RV
planet searches due to their high level of intrinsic RV variability. Magnetic activity in the form
of spots, plage, changes in the convection pattern, etc. can cause significant RV {\em jitter} of
several tens of m\,s$^{-1}$ (Saar \& Donahue 1997, Saar \& Fischer 2000, Hatzes 2002). This
additional source of noise can obscure the signal due to the presence of a planetary companion. In
spite of this large intrinsic noise planetary companions can be detected around young stars as long
as one has enough measurements to average out the activity noise (e.g. Hatzes et al. 2000). The TLS
Program may be the first RV planet search program to search for planets around a sample of young
stars and HD\,77407 is among the targets.

Using the cross-dispersed coud\'e echelle spectrograph, we obtained 32 RV measurements of HD\,77407
in 20 separate nights covering a time span of roughly two years. This spectrograph is especially
optimized for high-precision RV work, located in a temperature stabilized coud\'e room with a
temperature stabilized iodine I$_2$ gas absorption cell placed in front of the entrance slit. The
use of an I$_2$ cell allows modelling of temporal and spatial variations of the instrumental
point-spread function. Temperature sensors on the iodine cell, close to the grating, and close to
the grism allow us to monitor any possible change in temperature.  With the 1.2\,$''$ slit used the
resolving power $\lambda/\Delta \lambda$ is 67000. The cross-dispersing grism gives a wavelength
coverage of 4630 to 7370\,{\AA} which encompasses the full wavelength region of the iodine absorption
lines (5000 to 6000\,{\AA}). The RV analysis procedure largely follows the data modelling outlined by
Butler et al. (1996) and the instrumental point spread function reconstruction techniques of
Valenti et al. (1995). A template spectrum of the Tautenburg I$_2$ cell was taken at the McMath
telescope, and template stellar spectra taken without the cell were obtained with the TLS echelle
spectrograph.

\begin{figure} [htb] \resizebox{\hsize}{!}{\includegraphics[angle=0]{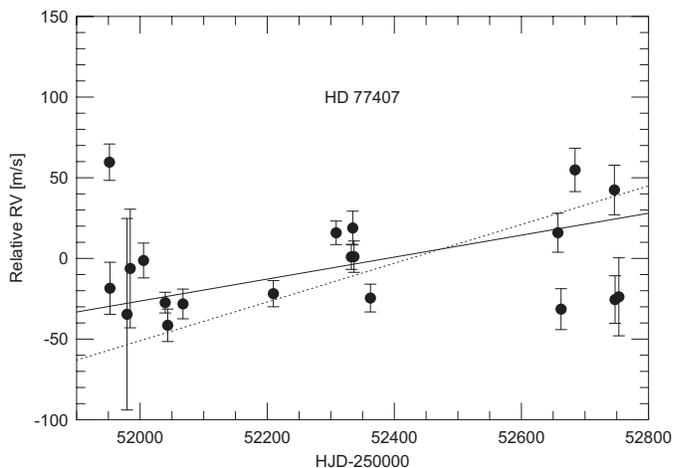}}
\caption{Radial velocity measurements of HD\,77407 (nightly averages) showing some scatter
consistent with an active spotted star. The lines show the expected RV change for an 0.3 and
0.6\,M$_{\odot}$ companion at 50\,AU separation. The lower mass is consistent with a long-term
trend seen in the RV data, and is also consistent with the imaging observations.}
\end{figure}

Fig.\,7 shows the RV measurements for HD\,77407 (nightly averages). The error bars represent {\em
internal} errors determined by the mean rms scatter of the RV computed using individual spectral
chunks (typically several hundred). This does not include any systematic errors which may be
present in the data and which are difficult to assess given the short time span of our
measurements. The rms scatter of the RV measurements in Fig.\,7 is about 30\,m/s, or more than a
factor of 2 greater than the internal errors. These variations could easily be caused by activity
noise. The Hipparcos photometry shows significant scatter of $\approx$ 0.015\,mag (after
subtracting the mean error in quadrature). This yields a spot filling factor of 1.4\,\% assuming
these are the source of the photometric variations. Hatzes (2002) presented an empirical
relationship between RV amplitude due to spots, the $v \cdot \sin i$ of the star, and the spot
filling factor. Assuming $v \cdot \sin i \simeq 7$\,km/s yields an RV amplitude from spots of about
80 m\,s$^{-1}$. Thus cool spots on HD\,77407 can easily account for the observed RV scatter of this
star. Our RV measurements, however, can exclude the presence of any companion with $m \cdot \sin i$
greater than $\sim 3$M$\rm_{Jup}$ and periods of less than 4 years. Such companions would cause
peak-to-peak variations in excess of 140\,m/s which would have been clearly visible in our data.

The acceleration induced by a 0.3\,M$_\odot$ and 0.6\,M$_\odot$ companion in a circular orbit
50\,AU from the star is 22.3 and 43.9\,m\,s$^{-1}$\,yr$^{-1}$, respectively. This expected motion
is shown in Fig.\,7 as solid (22.3\,m\,s$^{-1}$\,yr$^{-1}$) and dashed lines
(43.9\,m\,s$^{-1}$\,yr$^{-1}$). The acceleration expected from the lower mass companion is
consistent with any long term RV variations for HD\,77407 given the large scatter of the individual
measurements; such a lower-mass companion ($0.2$ to $0.4\,$M$_\odot$) at $\sim 50$\,AU projected
separation is detected in the imaging, (see above). Tentatively, the acceleration due to a higher
mass companion seems to produce too large of an RV change. However, we measure only the radial
component of total stellar orbital velocity, so we cannot exclude a higher mass for the stellar
companion. The long-term RV trend seen in HD\,77407\,A with an amplitude consistent with the mass
estimate for component B - as detected in the imaging - confirms that B is a bound companion to A.

Nidever et al. (2002) also included this star in their RV planet search program, but did not find
any variability within their time base of only 23 days consistent with our data.

\section{Discussion}

HD\,77407 is a slow rotating ($v \cdot \sin i \simeq 7$ km/s) G0 star with a notable chromospheric
excess emission in the H$\alpha$, H$\beta$ and Ca\,II lines. It is detected as radio source with a
flux of 1.67\,mJy at 20\,cm and also as EUV source (Wichmann et al. 2003). Montes et al. (2001a)
and Wichmann et al. (2003) report EW(Li\,I)=170...183\,m\AA, consistent with our own measurement
EW(Li\,I)=170\,m\AA, indicating that it is a very young star. It is identified as a member of the
local association (20 to 150\,Myrs) due to its galactic space motion (Montes et al. 2001a). Our RV
monitoring revealed no sub-stellar spectroscopic companion to HD\,77407, but some scatter probably
due to stellar activity (typical for young stars), and also a long-term trend consistent with the
mass estimate of the co-moving companion detected in the imaging.

\begin{figure*} [htb] \resizebox{\hsize}{!}{\includegraphics[angle=0]{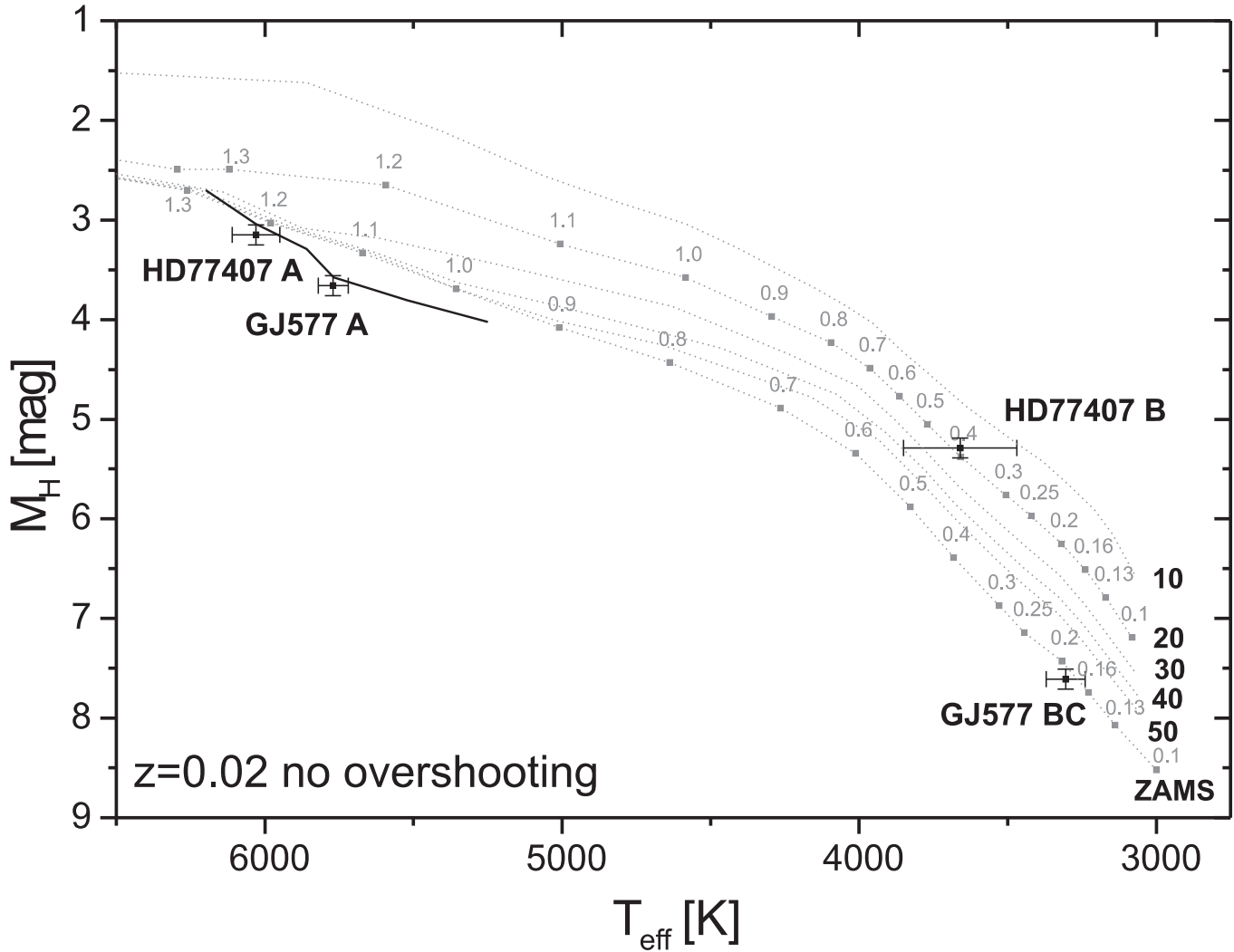}}
\caption{H-R diagram with Siess et al. (2000) models with (z=0.02 no overshooting) for HD\,77407
and GJ\,577 primary and secondary, our observed $M_{H}$ value versus the temperature obtained from
our observed (or derived) spectral type. The isochrones (dotted lines) are labeled by ages in Myrs.
The object mass is given in units of solar mass. Furthermore we have plotted data for main-sequence
stars (solid black line) from Schmidt-Kaler (M$_{V}$) and Kenyon-Hartmann (H-V and T$_{eff}$) for
spectral types F8V to K0V. GJ\,577\,A lies on that main-sequence for spectral type G5V as observed,
inconsistent with the ZAMS by Siess et al. by $\sim$500\,K.}
\end{figure*}

GJ\,577 is a G5\,V star with high levels of photospheric magnetic activity and a photometrically
determined rotation period of around 4 days (Messina \& Guinan 1998, Messina et al. 1999). Based on
the kinematic criteria this star can be considered as a member of the Hyades supercluster with an
age of 600\,Myrs. A moderate Ca\,II H and K emission is observed in the spectra as well as an
EW(Li\,I) of 145\,m\AA . This Lithium absorption is too strong for a member of the Hyades
supercluster and is close to the weakest Li-lines of the local association with an age of 20 to
150\,Myrs (Montes et al. 2001a, 2001b). Halbwachs et al. (2003) included GJ\,577 in their
spectroscopic observing program, but did not find any close low mass companion, nor any long-term
trend, consistent with our data.

If we use the measured absolute H-band magnitudes of HD\,77407\,B and GJ\,577\,B as well as the
stellar age given above, we can determine the companion mass and its effective temperature by using
the evolutionary models for low-mass stars from Siess et al. (2000). The spectral type is converted
to effective temperature using the scale for main-sequence dwarf stars by Kenyon \& Hartmann
(1995).

In Fig.\,8, we show the H-R diagram for both new visual pairs and compare their locations with
tracks and isochrones by Siess et al. (2000) with a metallicity z=0.02 and no overshooting.

GJ\,577\,B appears to be older than 100\,Myrs or already on the zero-age MS (ZAMS), also consistent
with the 20 to 150\,Myrs given by Montes et al. (2001a, 2001b). Here, it is not possible to
determine the age better from the companion alone, because the isochrones lie very close together
at its location in the H-R diagram.

We used Siess et al. (2000) models as well as Baraffe et al. (1998) models. We took those sets of
models which were available for low-mass objects down to 0.1\,M$_{\sun}$. For Baraffe et al. (1998)
models (mixing length parameter $\alpha=1.0$, He abundance $Y=0.275$ and solar metallicity [M/H]=0)
and for the Siess et al. (2000) models (with metallicity z=0.02 no overshooting and with
Kenyon-Hartmann conversion) the primaries lie below the main sequence, which is unphysical and may
indicate problems with that particular set of models. We note that the primary stars are not
saturated in the 2MASS images and that the Hipparcos parallaxes of the primaries should not be
affected by the much fainter companions. Furthermore we used M$_{V}$ data from Schmidt-Kaler and
converted them to M$_{H}$ using (V-H) from Kenyon-Hartmann (1995). Those data are consistent with
the 2MASS photometry and lie also under the Siess et al. ZAMS (see Fig.8). For a metallicity
[M/H]=-0.5, mixing length parameter $\alpha$=1 and He abundance Y=0.25 for Baraffe models and
metallicity z=0.01 for Siess models, each of the two pairs appears to be co-eval. However it is
extremely unlikely that these two young stars are that metal-poor. Hence we do not show those
models here.

GJ\,577\,B was also detected by McCarthy et al. (2001) and Lowrance et al. (2003). The former show
their J-band image having resolved A and B. They give $J = 11.15\pm0.15$\,mag and $I \simeq
13$\,mag, consistent with spectral type mid-M. Furthermore, McCarthy et al. (2001) quote Lowrance
(2001) as having measured the proper motion of GJ\,577\,B to be consistent with A. McCarthy et al.
(2001) do not show nor mention a spectrum, and the dissertation of Lowrance (2001) is not available
to us. Lowrance et al. (2003) show that GJ\,577\,B is actually a double star and is called
GJ\,577\,B\&C. The components B and C are both close to the sub-stellar limit or brown dwarfs. They
have a combined spectral type of M5 to M6 which is marginally consistent with our results.

We can use the ALFA and Keck images to obtain limits for undetected, but detectable faint companion
candidates, i.e. for determining the limiting dynamic range achieved in terms of magnitude
difference versus separation. We measure the $3 \sigma$ flux level (for HD\,77407 for both Keck and
ALFA, so that we can compare them), and ratio it to the primary to determine the curves shown in
Fig.\,9.

\begin{figure*} [htb] \resizebox{\hsize}{!}{\includegraphics[angle=0]{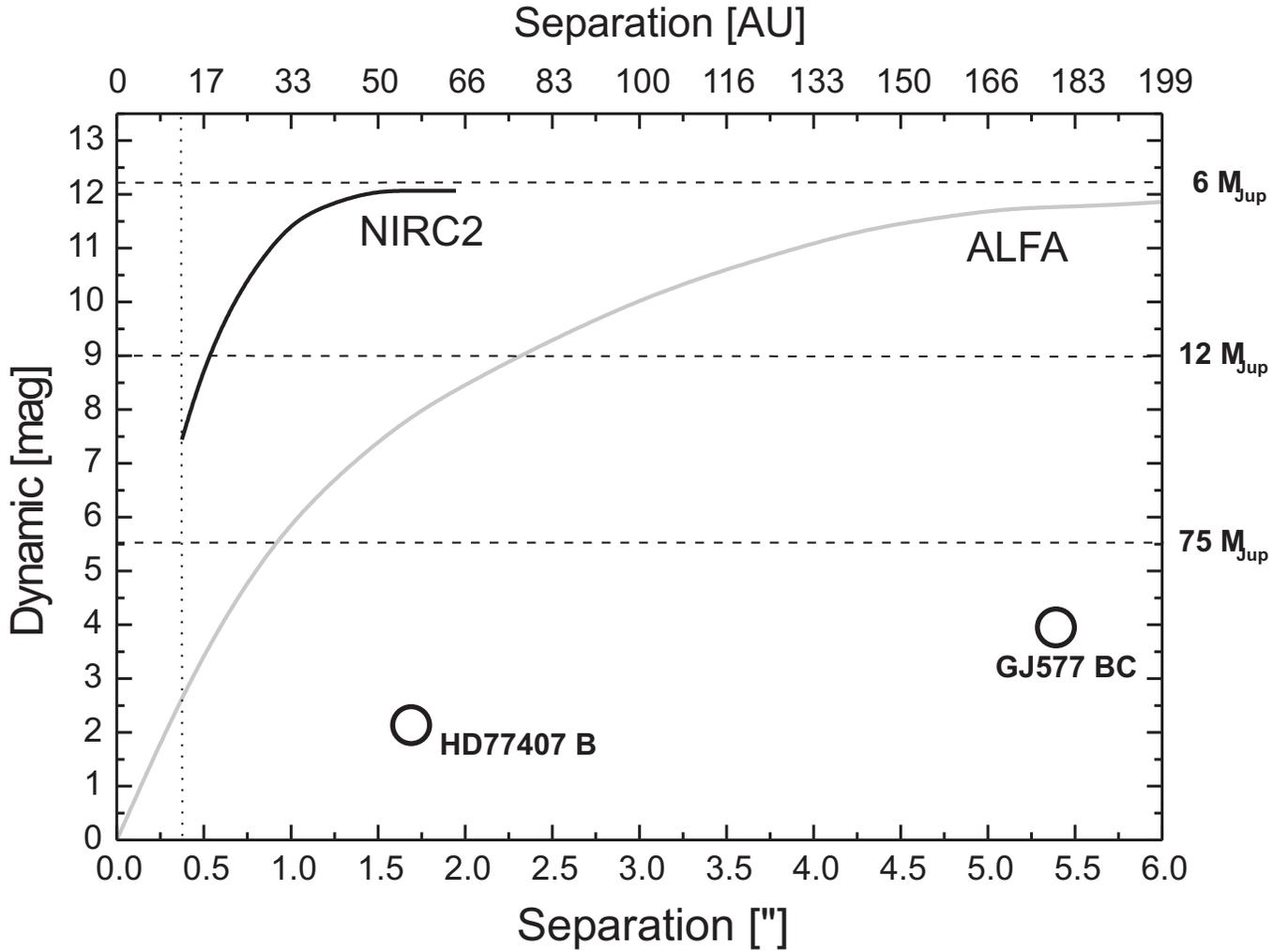}}
\caption{Dynamic ranges achieved with ALFA and Keck AO for HD\,77407, magnitude difference in H
between primary and companions versus separation in arcsec. The two companions detected are
indicated by circles. The Keck system is obviously much more sensitive than ALFA. We also indicate
magnitude differences expected for companions at the upper mass limits for planets and brown dwarfs
- computed for an age of 50\,Myrs using Baraffe et al. (2003). Any brown dwarf should have been
detected outside of 0.5\,arcsec (17\,AU), and any giant planet above $\sim 6.5$\,M$\rm_{Jup}$ would
have been detected outside of 1.5\,arcsec. The upper x-axis scale in AU is given for the distance
of HD\,77407.}
\end{figure*}

Any additional stellar companion (above $\sim 75$\,M$\rm_{Jup}$ for 50\,Myrs following Baraffe et
al. 2003) should have been detected at a separation of $\ge 0.4$\,arcsec, the radius of the
semi-transparent coronograph used. Also, any brown dwarf companion above $\sim 40$\,M$\rm_{Jup}$
would have been detected outside of 0.4\,arcsec. Brown dwarfs of any mass ranging from $\sim 12$ to
$\sim 75$\,M$\rm_{Jup}$ were detectable outside of 0.5\,arcsec, which is 17\,AU at the distance of
HD\,77407. We reached a magnitude difference of $\Delta H=9$\,mag at 0.5\,arcsec separation. Giant
planets with masses from $\sim 6.5$ to 12\,M$\rm_{Jup}$ were detectable outside of 1.5\,arcsec
(50\,AU), where we reached $\Delta H=12$\,mag.

We will report other detected companion candidates (both background objects as well as possibly
bound secondaries) as well as other possibly negative (null) results (i.e. primaries without any
detected companion candidates) elsewhere.

\begin{table} [htb]
\centering \caption{Stellar properties of HD\,77407\,B and GJ\,577\,B} \label{prop}
\begin{tabular}{l|l|l|l}
& HD\,77407\,B & GJ\,577\,B\\
\hline
Age $[$Myrs$]$    & 10 ... 40       & $>$ 100   \\
Mass [M$_{\sun}$] & 0.3 ... 0.5     & 0.16 ... 0.2  \\
Spec type           & M0 ... M3     & M4 ... M5 \\
T$_{\rm eff}$~$[K]$ & 3850 ... 3470 & 3370 ... 3240 \\
\end{tabular}
\end{table}

\acknowledgements {We are grateful to our referee, Eduardo Mart\'in, for several useful comments
and the Calar Alto Time Allocation committee for continuous support. RJ acknowledges support from
NASA Origins grant NAG5-11905. The German-Spanish Astronomical Centre Calar Alto is operated by the
Max-Planck-Institute for Astronomy, Heidelberg, jointly with the Spanish National Commission for
Astronomy. We would also like to thank the technical staff of the TLS in Tautenburg for all their
help and assistance in carrying out the observations. We would also like to thank Jens Woitas,
Sebastian Els, and Martin K\"urster for taking some of the spectra of HD\,77407. We made use of the
2MASS public data releases and of the Simbad database operated at the Observatoire Strassburg. Some
of the data presented herein were obtained at the W. M. Keck Observatory, which is operated as a
scientific partnership among the California Institute of Technology, the University of California
and the National Aeronautic and Space Administration. The authors would like to thank Randy Campbell and
David LeMignant for help during the observing run.}

{}

\end{document}